\documentclass[letterpaper,aps,prd,twocolumn,tightenlines,preprintnumbers,nofootinbib,showkeys,superscriptaddress]{revtex4-1}
\pdfoutput=1

\usepackage{XCharter}
\usepackage[T1]{fontenc}
\usepackage{mathptmx}

\usepackage{mathtools}
\usepackage{eqnarray}
\usepackage{fullpage}
\usepackage{amsfonts}
\usepackage{amsmath}
\usepackage{slashed}
\usepackage{amssymb}
\usepackage{graphicx}
\usepackage{epic}
\usepackage{eepic}
\usepackage{epsfig}
\usepackage{latexsym}
\usepackage[dvipsnames]{xcolor}
\usepackage[export]{adjustbox}
\usepackage{float}
\usepackage{multirow}
\usepackage{hyperref}
\usepackage{enumitem}
\hypersetup{colorlinks=true,citecolor=red,linkcolor=NavyBlue,urlcolor=NavyBlue}
\usepackage[caption=false]{subfig}
 
\usepackage{natbib}
\usepackage{relsize}
\usepackage[left=1.6cm,right=1.6cm,top=1.75cm,bottom=1.75cm]{geometry}
\usepackage{shorthand}
\begin{document}

\title{Testing left-right symmetry with an inverse seesaw mechanism at the LHC}

\author{Mathew Thomas Arun}
\email{mathewthomas@iisertvm.ac.in}
\affiliation{Indian Institute of Science Education and Research Thiruvananthapuram, Vithura, Kerala, 695 551, India} 

\author{Tanumoy Mandal}
\email{tanumoy@iisertvm.ac.in}
\affiliation{Indian Institute of Science Education and Research Thiruvananthapuram, Vithura, Kerala, 695 551, India}

\author{Subhadip Mitra}
\email{subhadip.mitra@iiit.ac.in}
\affiliation{Center for Computational Natural Sciences and Bioinformatics, International Institute of Information Technology, Hyderabad 500 032, India}

\author{Ananya Mukherjee}
\email{ananyatezpur@gmail.com}
\affiliation{Department of Physics, University of Calcutta, 92 Acharya Prafulla Chandra Road, Kolkata 700 009, India} 

\author{Lakshmi Priya}
\affiliation{Indian Institute of Science Education and Research Thiruvananthapuram, Vithura, Kerala, 695 551, India}
\affiliation{Deutsches Elektronen-Synchrotron DESY, Notkestr. 85, 22607 Hamburg, Germany}

\author{Adithya Sampath}
\affiliation{Indian Institute of Science Education and Research Thiruvananthapuram, Vithura, Kerala, 695 551, India}

\date{\today}

\begin{abstract}
In the left-right symmetric models, a heavy charged gauge boson $W'$ can decay to a lepton and a right-handed neutrino (RHN). If the neutrino masses are generated through the standard type-I seesaw mechanism, the Yukawa couplings controlling two-body decays of the RHN become very small. As a result, the RHN decays to another lepton and a pair of jets via an off-shell $W'$. This is the basis of the Keung-Senjanovi\'{c} (KS) process, which was originally proposed as a probe of lepton number violation at the LHC. However, if a different mechanism like the inverse seesaw generates the neutrino masses, a TeV-scale RHN can have large Yukawa couplings and hence dominantly decay to a lepton and a $W$ boson, leading to a kinematically different process from the KS one. We investigate the prospect of this unexplored process as a probe of the inverse seesaw mechanism in the left-right symmetric models at the High Luminosity LHC (HL-LHC). Our signal arises from the Drell-Yan production of a $W'$ and leads to two high-$p_T$ same-flavour-opposite-sign leptons and a boosted $W$-like fatjet in the final state. We find that a sequential $W'$ with mass up to $\sim 6$~TeV along with a TeV-scale RHN can be discovered at the HL-LHC. 
\end{abstract}

\maketitle

\section{\label{sec:level1}Introduction}\label{sec:intro}

\noindent
The neutrino oscillation data unambiguously establish that neutrinos have tiny but nonzero masses,
the explanation of which calls for physics beyond the Standard Model (SM).
Among the various extensions of the SM, the left-right symmetric models (LRSMs)~\cite{Senjanovic:1975rk,Mohapatra:1974gc,Mohapatra:1980qe,Senjanovic:1978ev,Pati:1974yy} provide a natural framework to embed the right-handed neutrinos (RHNs) and generate the neutrino masses. In this framework, parity violation is a low-energy feature; parity invariance is restored at high energies. Hence, the SM gauge group is extended to $SU(3)_C \otimes SU(2)_L \otimes SU(2)_R \otimes U(1)_{B-L}$ in the LRSM and the left-right symmetry breaks at the TeV scale introducing two heavy gauge bosons, $W^\prime$ and $Z^\prime$ in the spectrum.

A simple method to generate the small masses of the left-handed neutrinos is the seesaw mechanism,
where one introduces a set of heavy SM-singlet Majorana fermions breaking the $(B-L)$ symmetry.
However, the mechanism relies on a very high-scale explicit breaking of the lepton number symmetry in its simplest form (type-I seesaw~\cite{Minkowski:1977sc,Mohapatra:1979ia}). Hence, in this case, one requires the new physics scale $M$ to be of the order of $10^{14}$ GeV, i.e., much beyond the reach of colliders, to arrive at the observed neutrino mass scale $\displaystyle v^2/M\lesssim 0.1 $ eV without fine tuning the Higgs vacuum expectation value (VEV, $v$).
From a collider perspective, a very interesting possibility appears when one considers the inverse seesaw mechanism (ISM)~\cite{Mohapatra:1986aw,Mohapatra:1986bd}. In the ISM, one obtains the sub-eV neutrinos by adding three extra singlet Majorana fermions with mass $\mu \sim$ keV~\cite{Bazzocchi:2010dt,Dias:2011sq} with the three TeV-range RHNs. 

\begin{figure*}[t]
\captionsetup[subfigure]{labelformat=empty}
\subfloat[(a)]{\includegraphics[width=0.3\textwidth]{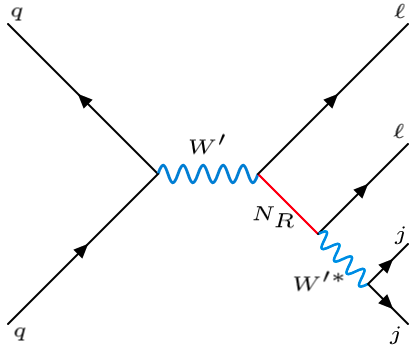}\label{fig:KS1}}\hfill
\subfloat[(b)]{\includegraphics[width=0.3\textwidth]{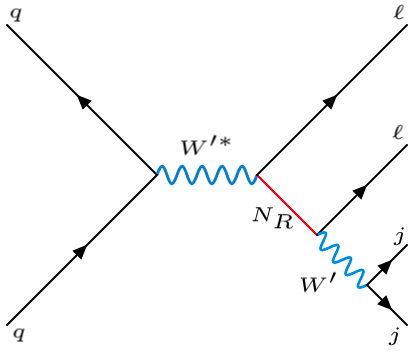}\label{fig:KS2}}\hfill
\subfloat[(c)]{\includegraphics[width=0.3\textwidth]{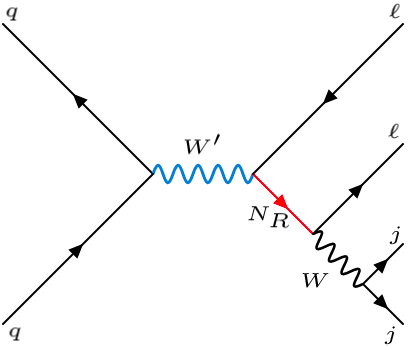}\label{fig:Signal}}
\caption{Feynman diagrams for the KS Process: (a) for $M_{W^\prime} > M_{N_R}$ and (b) for $M_{W^\prime} < M_{N_R}$. Our signal process with $M_{W^\prime} > M_{N_R}$ and the decay of $N_R$ through on-shell $W$ is shown in (c).}\label{fig:fynd}
\end{figure*} 

In this paper, we investigate an interesting collider signature of a minimal realisation of the LRSM with the ISM. We consider the $pp\to W^\prime$ process where $W^\prime$ is a heavy charged gauge boson which decays to a charged lepton ($\ell$) and a RHN ($N_R$). The heavy RHN then further decays to a $W$ boson and another charged lepton (of the same flavour; in general, the charged leptons can have different flavours, but we do not consider this possibility here). For this decay to occur, we need $M_{N_R} < M_{W^\prime}$, a mass ordering possible to obtain with the ISM. 
When the $W$, produced in the decay of $N_R$, decays hadronically, the process becomes similar to the Keung-Senjanovi\'{c} (KS) process~\cite{Keung:1983uu}, which was proposed as a
direct probe of the Majorana nature of the RHNs in the LRSM. The KS process also acts as a clean probe of the heavy right-handed $W'$ and, thus, the TeV-range breaking of the left-right symmetry. In the original KS process, the mass of the RHN was taken to be lighter than the mass
of $W'$, i.e., $M_{N_R}< M_{W^\prime}$. In
this process, a RHN, produced along with a charged lepton through the Drell-Yan production of $W^\prime$, decays to a charged lepton and two jets through an off-shell $W^\prime$, as shown in Fig.~\ref{fig:KS1}. If, however, $M_{W^\prime} < M_{N_{R}}$, a similar final state would arise, but in this case, $N_R$ would be produced through an off-shell $W^{\prime}$ and decay through an on-shell $W^\prime$, as shown in Fig.~\ref{fig:KS2}. Either way, we would get the same $\ell\ell jj$ final state in both cases~\cite{Ferrari:2000sp,Gninenko:2006br,Atre:2009rg,Nemevsek:2011hz,Chen:2011hc,Chakrabortty:2012pp,Aguilar-Saavedra:2012grq,Han:2012vk,Chen:2013foz,Rizzo:2014xma,Gluza:2015goa,Ng:2015hba,Dev:2015kca,Das:2016akd,Mitra:2016kov,Roitgrund:2020cge,Das:2017hmg,Arbelaez:2017zqq,Nemevsek:2018bbt}.

Kinematically, our signature differs from the KS process mainly in the fact that in our case the RHN decays to an on-shell $W$ [in particular, $N_R\to \ell W_h$, where $W_h$ denotes a hadronically decaying $W$, see Fig.~\ref{fig:Signal}] as opposed to a $W^\prime$ (off shell or on shell). In a regular LRSM with type-I seesaw, the Yukawa couplings responsible for the two-body decays, $N_R\to \ell W$, are extremely small for TeV-scale RHNs. As a result, such decays become negligible compared to off-shell $W'$-mediated three-body decay. However, with the ISM, the Yukawa couplings that govern the $N_R\to \ell W$ decay can be sufficiently large such that the signature is observable at the LHC. In the original KS process, the two leptons in the final state have the same charge half of the time because of the Majorana nature of the $N_R$, whereas, as we shall see, they always have opposite charges in our model. Even if we ignore the charges of the lepton pair (as is often done in the experimental searches), making the final state of the process essentially the same as that of the KS one, the kinematics of these two processes are very different. In our signal, the jet pairs come from the decay of boosted $W$ and thus form a $W$-like fatjet. This feature is absent in the KS process. Hence, the two processes complement each other and can be used to probe different neutrino mass generation mechanisms. Unlike the KS process, the $N_R\to \ell W_h$ signature has not been searched for at the LHC.

The LHC experiments are yet to search for the particular signal we consider, but similar signatures were considered earlier in some phenomenological studies. In Ref.~\cite{Nemevsek:2012iq}, a rare decay of $N_R$ to the left-handed leptons through a $W$ boson is considered as a probe for the heavy-light neutrino mixing angles. Ref.~\cite{Han:2012vk} looks at the opposite-sign dilepton plus jets signature as a probe of the chiral couplings of $W'$. Limits on the heavy-light neutrino mixing, lepton-flavour-violating processes, and their LHC prospects have been studied using the $\ell\ell jj$ channel in Refs.~\cite{Chen:2013foz, Lee:2013htl}. The sensitivity of the future high-energy LHC to the Dirac Yukawa coupling of the heavy neutrino in the LRSM with the ISM is available in Ref.~\cite{Helo:2018rll}. The $\ell\ell jj$ channel considered in the above references forms an important probe of the heavy-light neutrino mixing. It also acts as a test of the existence of $W'$ and $N_R$ together. We investigate the prospects of this important channel at the $14$ TeV High Luminosity LHC (HL-LHC) with the modern jet-substructure technique.

The paper is organised as follows. In Section~\ref{sec:MLRSMISS}, we discuss the model and neutrino parameters, in Section~\ref{sec:compKS}, we compare our signature and the KS process in more detail, in Section~\ref{sec:exclu}, we discuss the LHC bounds on $W^\prime$. In Section~\ref{sec:lhcpheno}, we analyse our signature. Finally, in Section~\ref{sec:sumcon}, we conclude.

\begin{table}[!t]
\centering{
\begin{tabular*}{\columnwidth}{l @{\extracolsep{\fill}} cccc }
\hline
 Particle &  $SU(3)_C$ &  $SU(2)_L$ & $SU(2)_R$ & $U(1)_{B-L}$\\ \hline\hline
$q_L^i \equiv \begin{pmatrix} u_L^i \\ d_L^i \end{pmatrix}$ & $\mathbf{3}$ & $\mathbf{2}$ & $\mathbf{1}$ & $\dfrac{1}{3}$ \\
$q_R^i \equiv \begin{pmatrix} u_R^i \\ d_R^i \end{pmatrix}$  & $\mathbf{3}$ & $\mathbf{1}$ & $\mathbf{2}$ & $\dfrac{1}{3}$ \\ \hline
$L_L \equiv \begin{pmatrix} \nu_L^i \\ e_L^i \end{pmatrix}$ & $\mathbf{1}$ & $\mathbf{2}$ &$\mathbf{1}$ & $-1$ \\
$L_R \equiv \begin{pmatrix} N_R^i \\ e_R^i \end{pmatrix}$ & $\mathbf{1}$ & $\mathbf{1}$ &$\mathbf{2}$ & $-1$ \\  \hline
$S^i$ & $\mathbf{1}$ & $\mathbf{1}$ & $\mathbf{1}$ & $0$ \\ \hline
$\Phi = \left(\begin{matrix}\phi^0_1 & \phi^+_2\\\phi^-_1 & \phi^0_2\end{matrix}\right)$ & $\mathbf{1}$ & $\mathbf{2}$ & $\mathbf{2}$ & 0 \\

$\chi_{L} = \left(\begin{matrix} \chi^+_L\\ \chi^0_L\end{matrix}\right)$ & $\mathbf{1}$ & $\mathbf{2}$ & $\mathbf{1}$ & $1$ \\ 

$\chi_{R} = \left(\begin{matrix} \chi^+_R\\ \chi^0_R\end{matrix}\right)$ & $\mathbf{1}$ & $\mathbf{1}$ & $\mathbf{2}$ & $1$ \\ \hline
\end{tabular*}
\caption{Particle content of left-right symmetric model based on the gauge group $SU(3)_C \otimes SU(2)_L \otimes SU(2)_{R} \otimes U(1)_{B-L}$. }
\label{tab:partqn}}
\end{table}

\section{Left-right symmetry with inverse seesaw}
\label{sec:MLRSMISS}

\noindent
We consider a simple realisation of the LRSM with the following gauge structure,
\begin{equation}
\mc{G}_{LRSM} = SU(3)_C\otimes SU(2)_L\otimes SU(2)_R \otimes U(1)_{B-L}.
\label{LRgauge}
\end{equation}
In this model, the $SU(2)_R$ gauge coupling $g_R$ is a
free parameter, not same as the $SU(2)_L$ gauge coupling $g_L$.
The particle content and the gauge quantum numbers are summarised in Table~\ref{tab:partqn}.
In addition to the SM fermions, there are three right-handed neutrinos ($N_R^i$) and three neutral fermions ($S^i$) introduced for the three generations. The RHNs are naturally present in the LRSM, whereas the neutral fermions are required to realise the ISM. The RHNs form the $SU(2)_R$ doublets with the right-handed charged leptons,  whereas the $S^i$ are all singlets under $\mc{G}_{LRSM}$. The enlarged scalar sector consists of three multiplets, i.e., the Higgs bidoublet $\Phi$ and two doublet scalars $\chi_{L}$ and $\chi_{R}$, which are doublets under  $SU(2)_{L}$ and $SU(2)_R$ gauge groups,  respectively.

\subsection{Symmetry breaking}
\noindent
The spontaneous breaking of the gauge symmetry follows the pattern below:
\begin{align}
&~SU(3)_C\otimes SU(2)_L\otimes SU(2)_R \otimes U(1)_{B-L} \nn \\
&\hspace{2.6cm}~\Big\downarrow \langle \chi_R\rangle \nn \\
&\hspace{0.8cm}~SU(3)_C\otimes SU(2)_L\otimes U(1)_Y \nn \\
&\hspace{2.6cm}~\Big\downarrow \sqrt{\langle \Phi\rangle^2 + \langle \chi_L \rangle^2} \nn \\
&\hspace{1.45cm}~SU(3)_C\otimes U(1)_{EM}.\nn
\end{align}
The doublet field $\chi_R$ is nontrivially charged under $SU(2)_R \times U(1)_{B-L}$. Hence, when it acquires a TeV-scale VEV, the group breaks down to $U(1)_Y$ and gives masses to the charged $W^\prime$ and neutral $Z^\prime$ bosons. 
The standard electroweak symmetry breaking is then carried out by the neutral scalar fields in $\chi_L$ and the bidoublet $\Phi$ satisfying the relation
\begin{equation}
\sqrt{\Big(\left\langle\chi_L^0\right\rangle^2+ \left\langle\phi_1^0\right\rangle^2+ \left\langle\phi_2^0\right\rangle^2\Big)} \sim 246 \ \text{GeV}.
\end{equation}
The electromagnetic charge can be expressed as 
\begin{equation}
Q_{EM} = I_{3 L} + I_{3 R} + \dfrac{B-L}{2}.
\end{equation}

\begin{figure}
\centering
\includegraphics[width=\columnwidth]{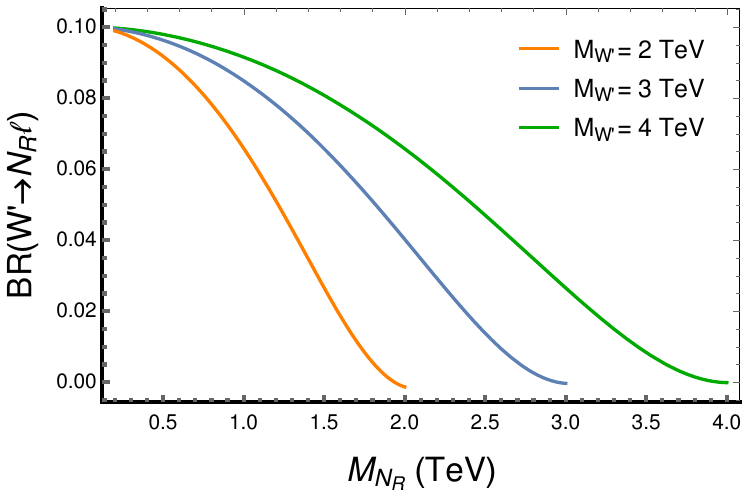}
\caption{Branching ratio of $W'\to N_R\ell$ as functions of $M_{N_R}$ for different $M_{W'}$ choices.}
\label{fig:BRWpNR}
\end{figure}

Since here our motivation is to study the unexplored signature as discussed in the Introduction, we make an assumption that the $W^\prime$ boson dominantly decays to the SM fermions and RHNs. We make a further simplifying assumption that two of the RHNs are heavier than  $W^\prime$, so that the vector boson can decay to only one generation of RHNs. Since $W^\prime$ universally couples to  right-handed fermions with the strength $g_R$, the branching ratio (BR) of the $W^\prime\to N_R\ell$ decay is about $10\%$ in the $M_{W^\prime}\gg M_{N_R}$ limit, as seen in Fig.~\ref{fig:BRWpNR}. As the mass of the RHN goes close to $M_{W^\prime}$, the BR of the $N_R\ell$ mode falls due to phase-space reduction.

\begin{figure*}
\captionsetup[subfigure]{labelformat=empty}
\subfloat[(a)]{\includegraphics[width=0.325\textwidth]{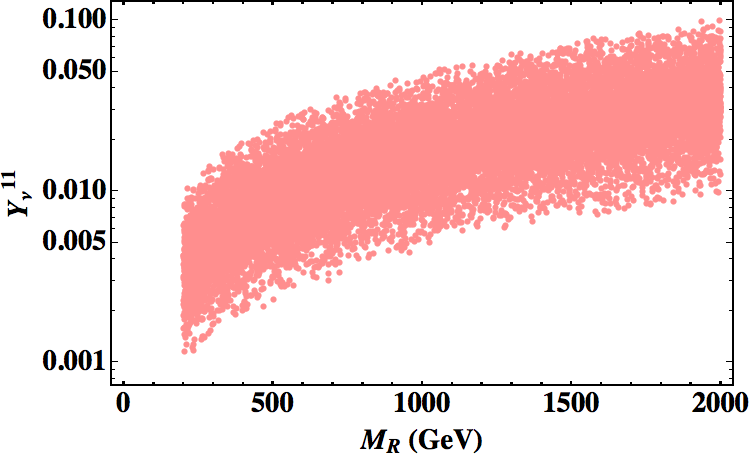}\label{fig:mr1}}\hfill
\subfloat[(b)]{\includegraphics[width=0.325\textwidth]{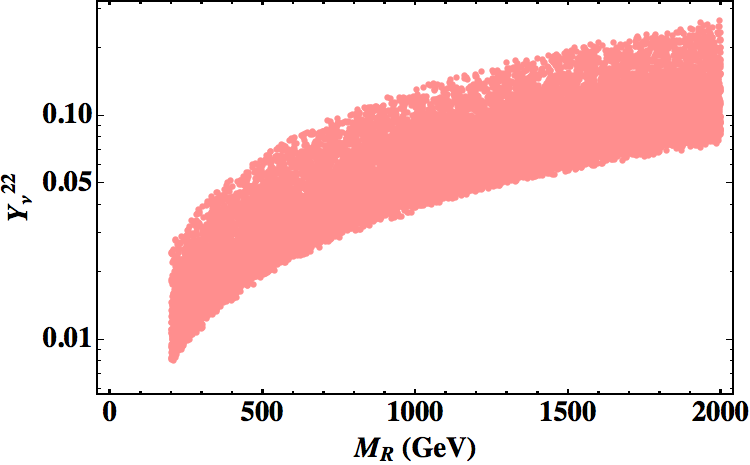}\label{fig:mr2}}\hfill
\subfloat[(c)]{\includegraphics[width=0.325\textwidth]{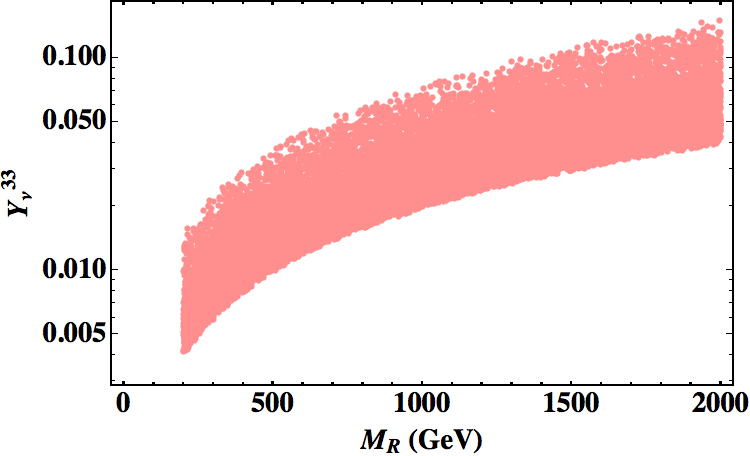}\label{fig:mr3}}\\
\subfloat[(d)]{\includegraphics[width=0.325\textwidth]{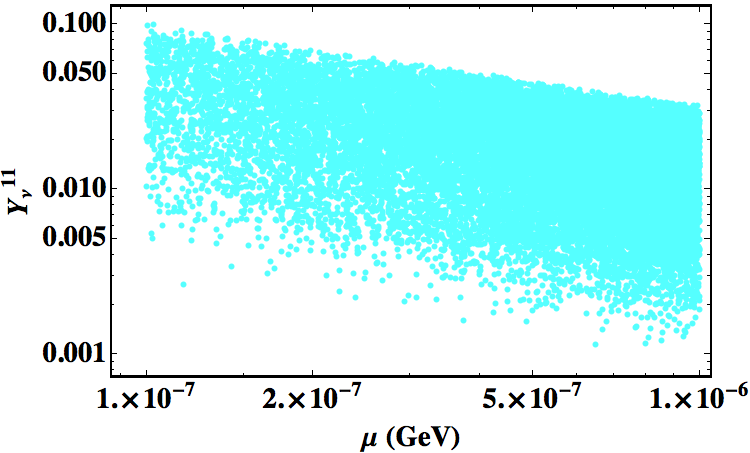}\label{fig:mu1}}\hfill
\subfloat[(e)]{\includegraphics[width=0.325\textwidth]{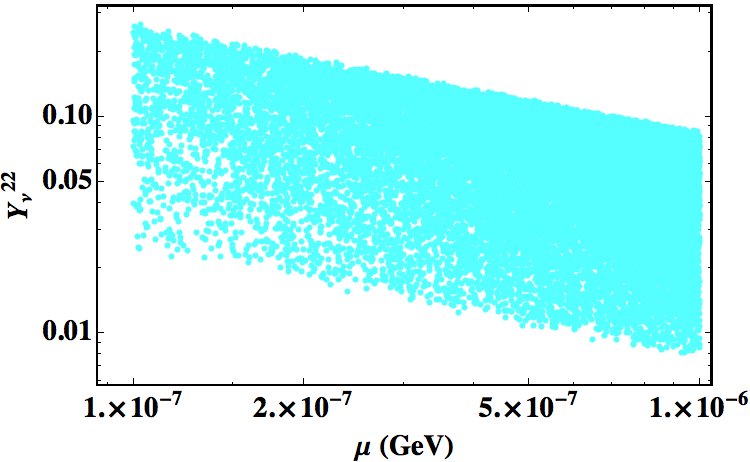}\label{fig:mu2}}\hfill
\subfloat[(f)]{\includegraphics[width=0.325\textwidth]{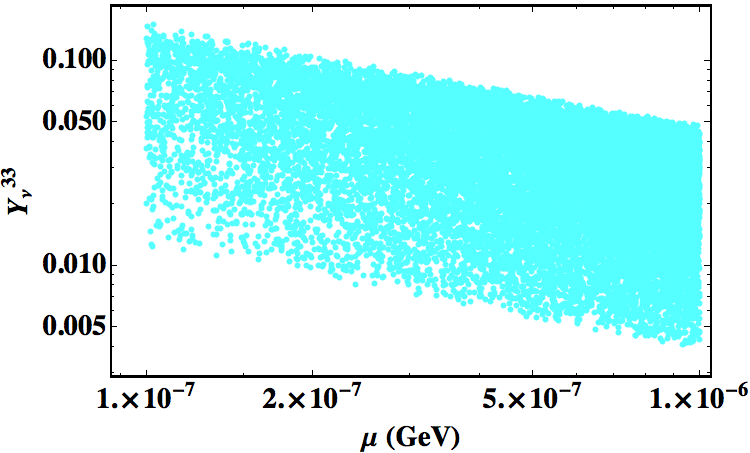}\label{fig:mu3}}
\caption{The Yukawa couplings vs RHN mass (top row) and the 
lepton-number-violating scale $\mu$ (bottom row) in the inverse seesaw scheme}\label{fig:iss_yuk}
\end{figure*}

\subsection{Inverse seesaw mechanism}
\label{sec:ISS}
\noindent
In the original ISM, three TeV-scale RHNs and three extra singlet neutral fermions $S_{L}^i$ (where $i\in \{1,2,3\}$) are added to the three active neutrinos $\nu_{L}^i$. The tiny neutrino masses are generated from the mixing between the $N_{R}^i$ and $S_{L}^i$ states,
\begin{equation}\label{ISS_Lag}
\mathcal{L} = Y \,\overline{L}_L \,\widetilde{H} \,N_{R} + M_{R} \,\overline{(N_{R})^{c}} \,S_{L}^c + \frac{1}{2} \mu \bar{S_{L}} (S_L)^c + {\rm H.c.},
\end{equation}
where $\widetilde{H} = i \sigma_2 H^*$, the superscript $c$ denotes charge conjugation, and we have suppressed the generation index.

In the left-right symmetric realisation of the ISM, 
the lepton masses are generated via a Yukawa Lagrangian,
\begin{align}
\mathcal{L}_{Y} =& - Y \bar{L}_R \Phi^\dagger L_L -\tilde{Y}\bar{L}_R \tilde{\Phi}^\dagger L_L - Y_1 \bar{S} \tilde{\chi}_L^\dagger L_L - Y_1 \bar{S}^c \tilde{\chi}_R^\dagger L_R \nn\\
&- \frac{1}{2} \mu \bar{S}^c S + {\rm H.c.},
\label{ISS_LR_Lag}
\end{align}
where $Y, \tilde{Y}, Y_1$ are the $3\times 3$ Yukawa couplings and $\mu$ is the $3\times 3$ Majorana mass matrix. In the above Lagrangian, $\tilde{\chi}$ and $\tilde{\Phi}$ represent the charge-conjugated fields.
The complete mass matrix  in the basis $( \nu_L, N_R^c, S^c)$ obtained from Eq.~\eqref{ISS_LR_Lag} reads as
\begin{equation}
M_\nu= \begin{pmatrix}\label{ISS_matrix}
0 && m^T_D && m'^T_D \\
m_D^T && 0 && M^T_R \\
m'_D && M_R && \mu \\
\end{pmatrix},
\end{equation}  
where
\begin{equation}
\begin{aligned}
m_D &= \frac{1}{\sqrt{2}} \Big(Y \langle\phi_1^0\rangle + \tilde{Y} \langle\phi_2^0\rangle \Big), & m'_D = \frac{1}{\sqrt{2}}Y_L \langle\chi_L^0\rangle, \\
M_R &= \frac{1}{\sqrt{2}}Y_R \langle\chi_R^0\rangle. \nonumber
\end{aligned}
\end{equation}
The light-neutrino mass can be found by a block diagonalisation of the mass matrix in the limit $\langle \phi_1^0 \rangle \gg \langle \phi_2^0 \rangle$ as
\begin{equation}
m_\nu \sim \frac{\langle \chi_L^0 \rangle}{\langle \chi_R^0 \rangle}\Big( m_D + m_D^T \Big) - m_D (M_R^T)^{-1}\mu M_R^{-1}m_D^T \ .
\end{equation}
The first term is the linear seesaw contribution and a consequence of the left-right symmetry. The experimentally observed neutrino mass squared differences and mixing angles predict this term to be subdominant, $\langle \chi_L^0 \rangle/\langle \chi_R^0 \rangle \lesssim 10^{-12}$. Even if $\chi_L^0$ is generated radiatively at one loop, it remains small and satisfies the above condition for $\mu \sim \mathcal{O}(1)$ keV and $ \langle\chi_R \rangle \sim 10^4$ GeV~\cite{Brdar:2018sbk}.

The novelty of the ISM with a TeV scale RHN lies in the double suppression by the mass scale associated with the mass scale $M$. 
To have a sub-eV neutrino mass scale with $m_D$ at the electroweak scale, one should have $M$ at the TeV scale and $\mu$ at the  keV scale. The tiny $\mu$ ensures the  neutrino mass to be small. As $\mu \rightarrow 0$, the lepton number symmetry is restored, leading to $m_\nu \rightarrow 0$. 
The neutrino Yukawa couplings can be obtained using the extended Casas-Ibarra formalism derived in Ref.~\cite{Dolan:2018qpy}, based on the original formalism~\cite{Casas:2001sr},
\begin{equation}\label{eq:CI}
Y_{\nu} = \frac{1}{v} \,U \,m_{\n}^{1/2} \,R \, \mu^{-1/2}\,M_{R}^T\,,
\end{equation}
with the matrices $m_\n = \text{diag}(m_1,m_2,m_3)$ and $M_R = \text{diag}(M_{R_1},M_{R_2},M_{R_3})$, carrying the light-neutrino mass eigenvalues and the heavy Majorana neutrino masses, respectively. Here, $R$ is a complex orthogonal matrix in general, and $U$ is the Pontecorvo-Maki-Nakagawa-Sakata matrix characterising the mixing among leptons.\footnote{For some recent implications of this kind of formalism, readers may see Refs.~\cite{Mukherjee:2021hed,Konar:2020wvl,Konar:2020vuu}.}

We numerically evaluate the Yukawa couplings using Eq.~\eqref{eq:CI}. We use the best-fit central values of all the oscillation parameters tabulated in Ref.~\cite{deSalas:2020pgw}. A flavour-symmetric realisation of such a model based on the left-right symmetry would restrict the neutrino oscillation parameters~\cite{Mukherjee:2015axj,Mukherjee:2017pzq}. We follow Ref.~\cite{Casas:2001sr} for the parametrisation of the rotational matrix ($R$). In general, $R$ can take any complex value for the mixing angles. However, we assume that the three angles in the rotational matrix are real for simplicity and choose them as $x= \pi/3$, $y= \pi/4$, and $z= \pi/5$ for a benchmark set of values. There is, however, nothing special about this particular set of values as our phenomenological analysis is mostly insensitive to the choice of these parameters.

In Fig.~\ref{fig:iss_yuk}, we show the dependence of the Yukawa couplings on the heavy RHN mass scale $M_R$ (top row) and the lepton-number-violating scale $\mu$ (bottom row). We see  that the Yukawa couplings increase with $M_R$ and decrease with $\mu$ from this figure [and also Eq.~\eqref{eq:CI}]. The order of magnitude of $\mu$ has a very crucial role in determining the degree of degeneracy among the RHN mass eigenstates obtained in the ISM. The keV-scale $\mu$ also plays an important role in resonant leptogenesis~\cite{Blanchet:2010kw}. The number of the same-sign vs opposite-sign dilepton events will be controlled by the $\mu$ parameter. In the limit $\mu\to 0$, the heavy neutrino becomes purely Dirac type, and the lepton number is conserved. In our case, a keV-scale 
$\mu$ is too small to produce any significant amount of same-sign events.

The BRs of the two-body decay modes of the RHNs, namely, $W^\pm\ell$, $Z\nu$, and $H\nu$, are determined by the Yukawa couplings shown in Fig.~\ref{fig:iss_yuk}. There is also a three-body decay mode of $N_R$ through an off-shell $W^\prime$ present in the LRSM. As explained before, the three-body decay is the dominant mode in the standard type-I seesaw where the Yukawa couplings controlling the two-body decays are very small. In that case, the KS process topology becomes important. With the ISM, the Yukawa couplings become large and, hence the two-body decays of $N_R$ through on-shell $W$ take over the three-body decays. Usually, the BRs of $N_R$ to $W\ell$, $Z\nu$ and $H\nu$ are in the proportion $2:1:1$ when the mass of $N_R$ is sufficiently above the kinematic threshold of these decays. This proportion can alter a bit if we fit neutrino data. In our collider analysis, we have taken the BRs of $N_R$ in the $2:1:1$ proportion for simplicity. However, the plots in Fig.~\ref{fig:iss_yuk} have been obtained by fitting latest neutrino data. The pair production of RHNs through an $s$-channel $Z'$ is also possible in our model [some discussion on the prospects of that channel can be found in Refs.~\cite{Das:2017flq,Das:2017deo,Choudhury:2020cpm,Deka:2021koh,Arun:2022ecj} in the context of $U(1)$ extended models]. 
Other interesting signatures of the LRSM with the ISM may appear in different sectors~\cite{Ezzat:2021bzs}.
Our signal is insensitive to the actual values of the Yukawa couplings as it appears in the BRs of $N_R$. This can be probed with high-precision in an electron-positron collider~\cite{Banerjee:2015gca}. 

\section{Signal topology and the KS process}
\label{sec:compKS}
\begin{figure*}[!t]
\captionsetup[subfigure]{labelformat=empty}
\subfloat[(a)]{\includegraphics[height=6.5cm,width=8.5cm]{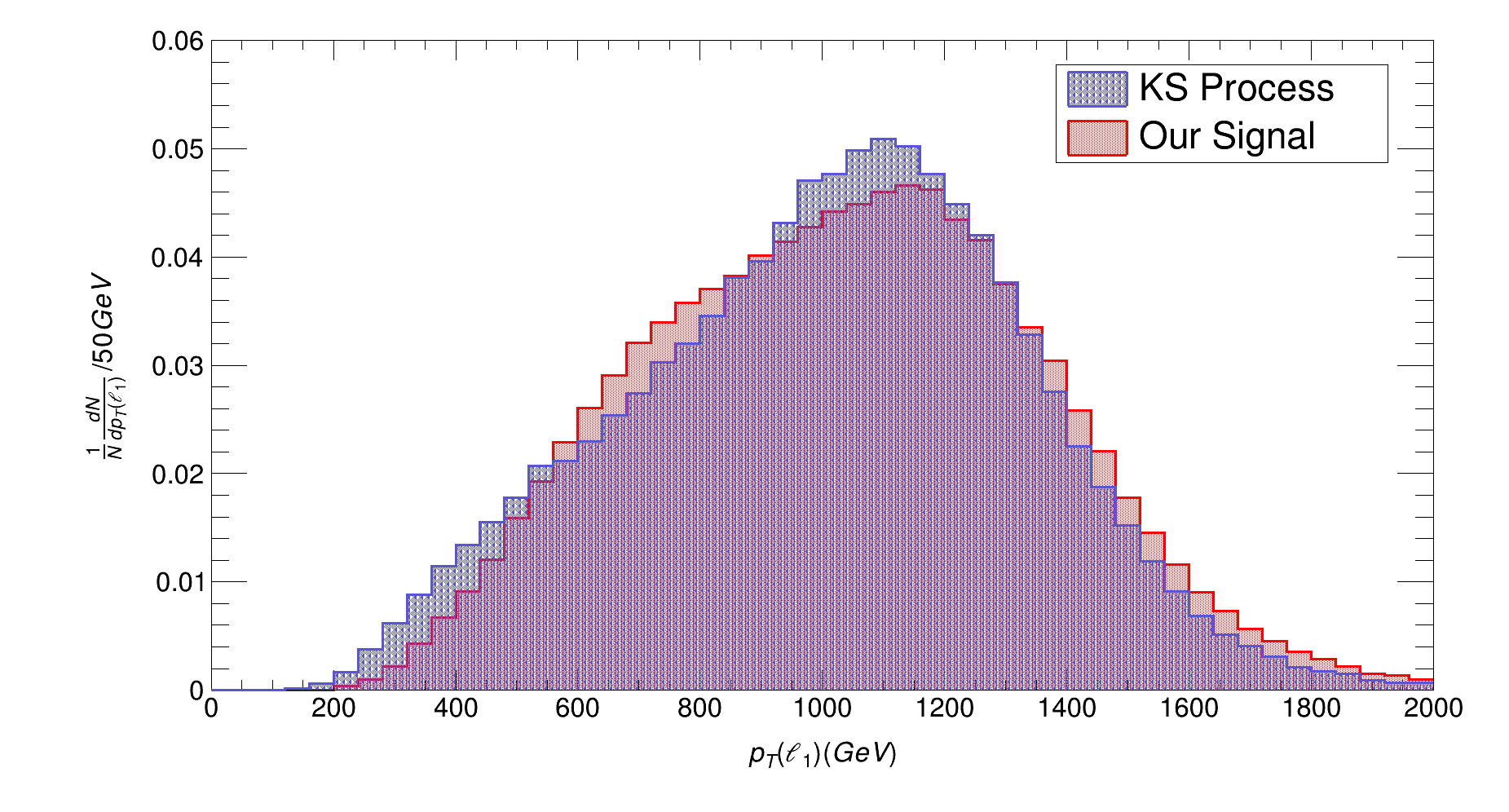}\label{fig:1stelpt}}\quad
\subfloat[(b)]{\includegraphics[height=6.5cm,width=8.5cm]{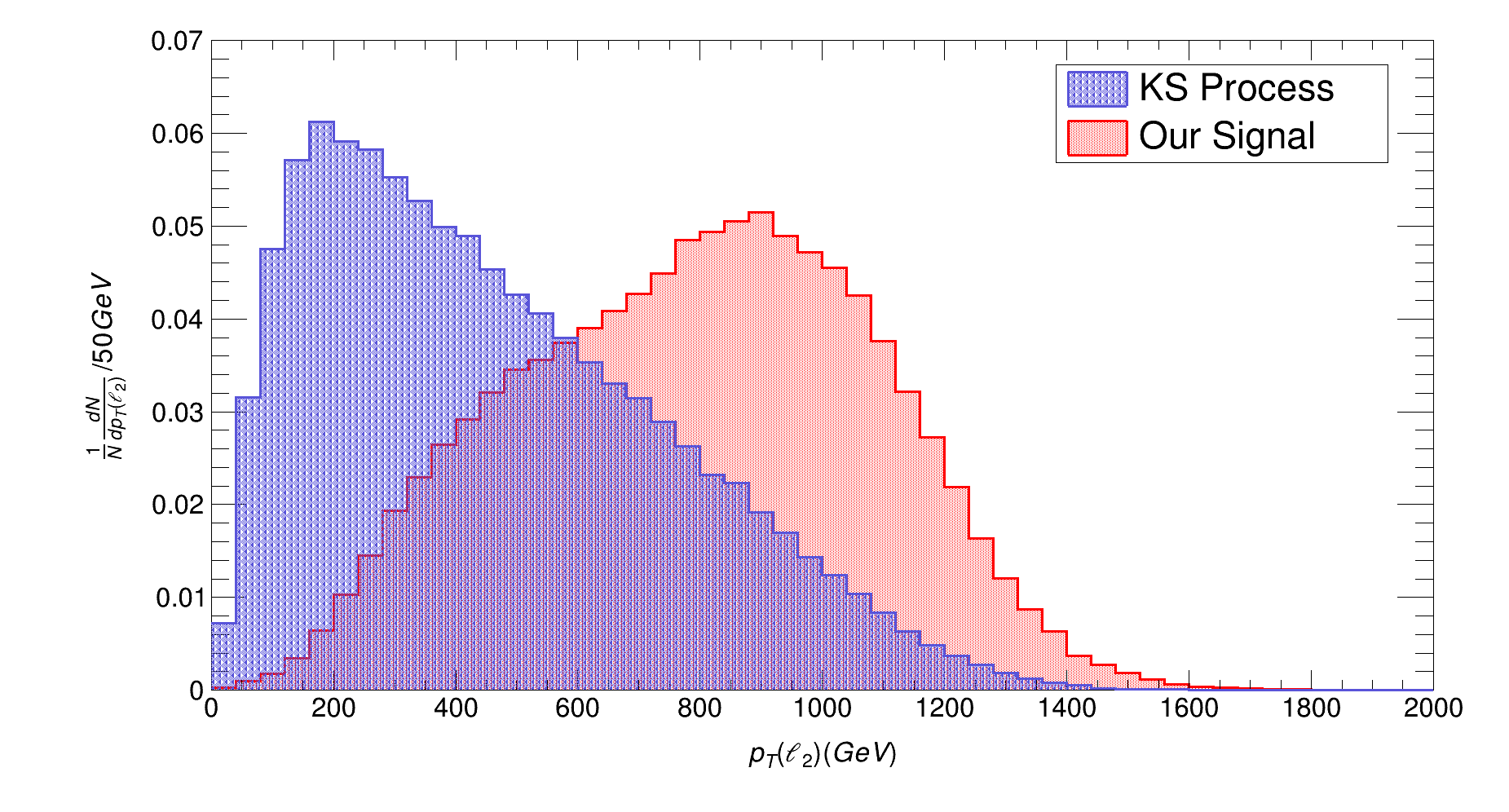}\label{fig:2ndelpt}}\\
\subfloat[(c)]{\includegraphics[height=6.5cm,width=8.5cm]{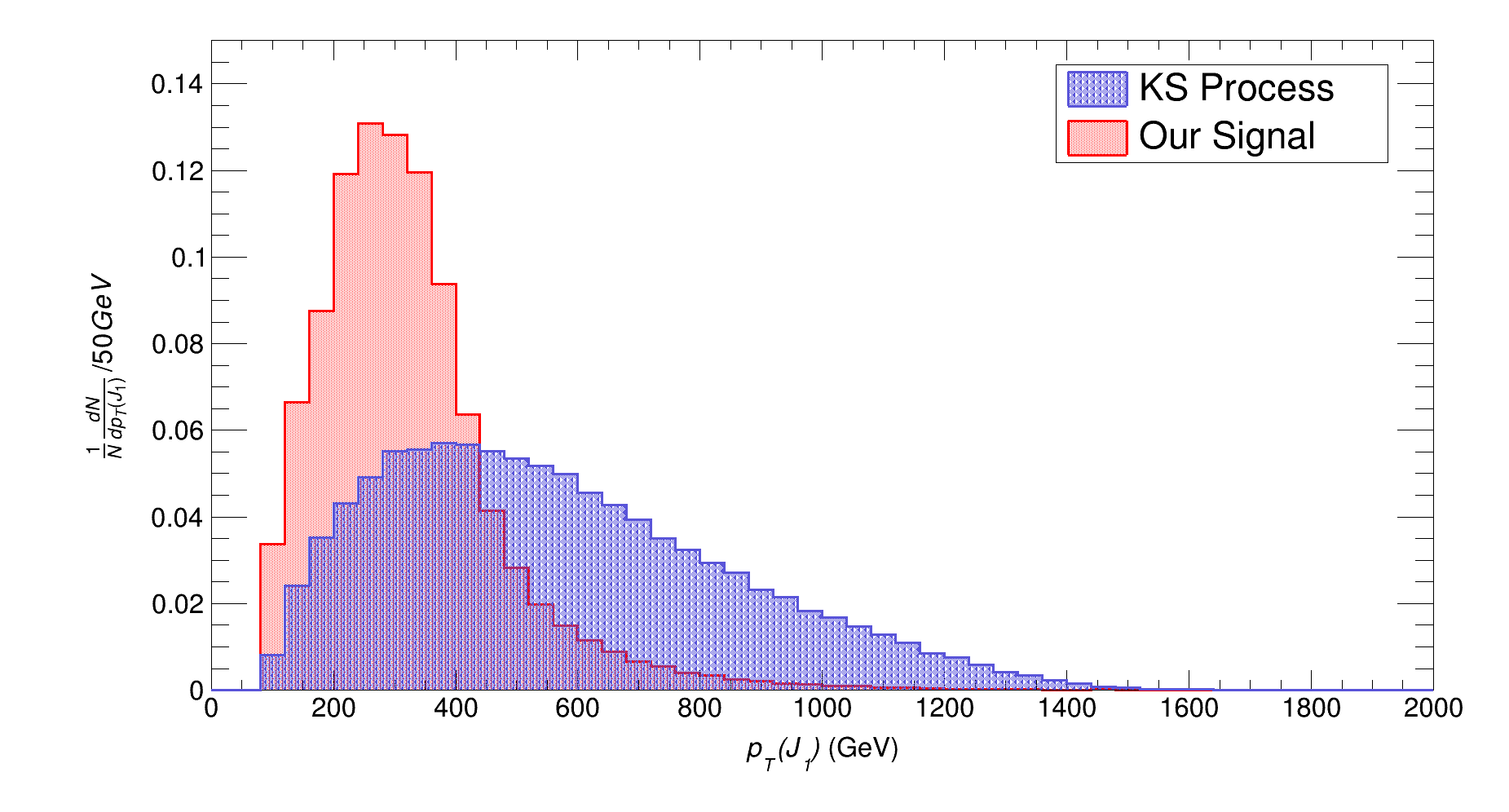}\label{fig:1stjtpt}}\quad
\subfloat[(d)]{\includegraphics[height=6.5cm,width=8.5cm]{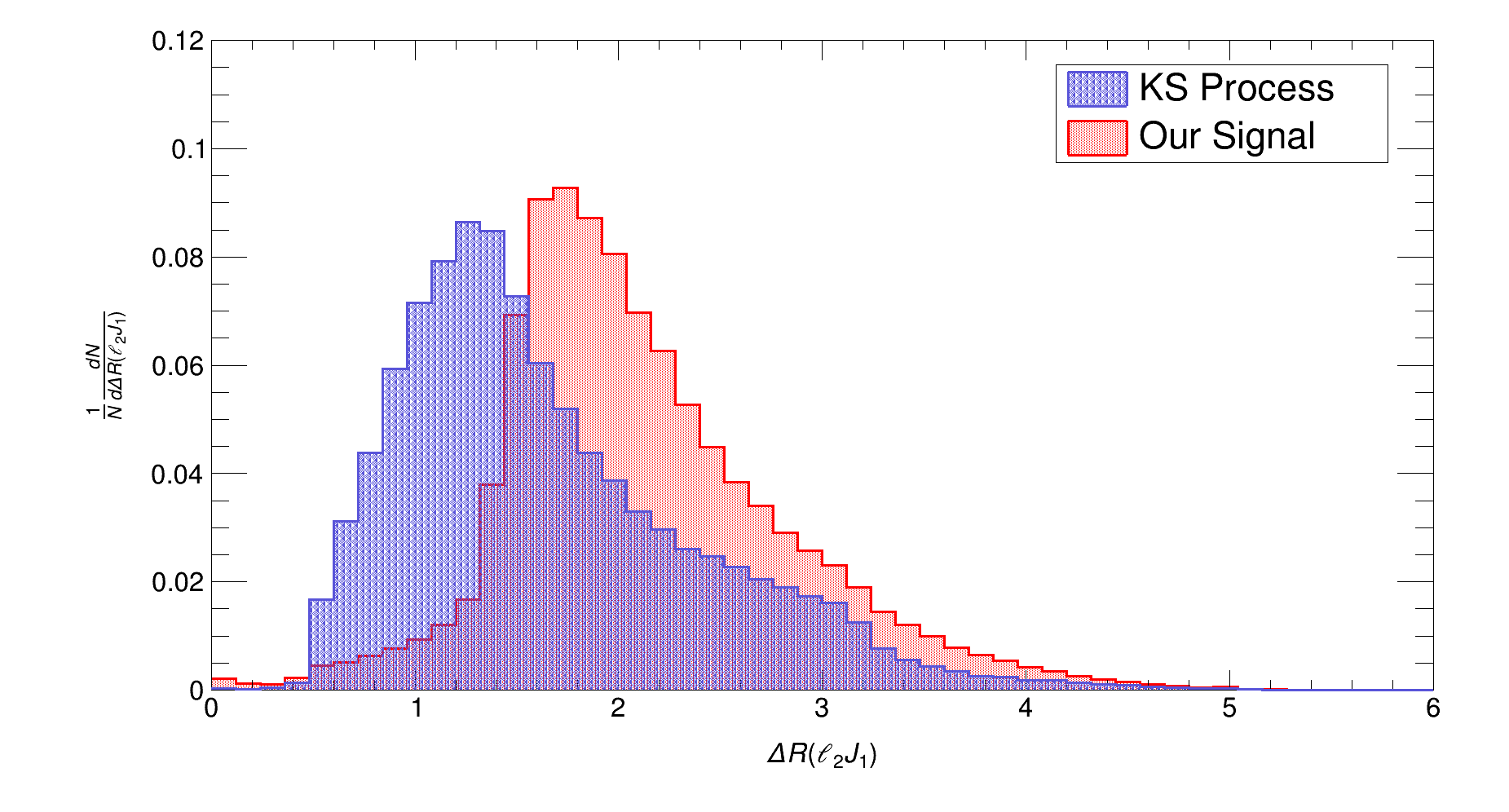}\label{fig:dlRe2j1}}
\caption{Comparison of various kinematic distributions of the KS process and our signal. These distributions are obtained for $M_{W'}=3$~TeV and $M_{N_R}=1$~TeV benchmark masses with $g_R=0.1$.}\label{fig:dist}
\end{figure*} 

\noindent
The KS process gives rise to the $\ell\ell jj$ final state. To be specific, the leptons are same flavour and same sign in nature. The same process ignoring the charges of leptons has been searched for by both the ATLAS and the CMS collaborations (see, e.g., Refs.~\cite{ATLAS:2018dcj,CMS:2021qef}). When $M_{W^\prime} < M_{N_R}$, the process is kinematically suppressed and hence, difficult to probe. The $M_{W^\prime} \gtrsim M_{N_R}$ region is accessible and can be further categorised into two kinematic regions---resolved and merged~\cite{Nemevsek:2018bbt}: 

\begin{enumerate}

    \item[(a)] When the RHN is not much lighter than $W^\prime$, i.e.,  $0.1\, M_{W^\prime}\lesssim M_{N_R} < M_{W^\prime}$, the two jets from the $W^{\prime*}$ decay can be resolved. This leads to two isolated leptons and at least two high $p_T$ jets. This process is fully reconstructible: the $jj$ system along with one lepton can be used to reconstruct the $N_R$. The invariant mass of the $\ell\ell j j$ system forms a peak around $W'$ mass.
    
    \item[(b)] In the merged topology,  $M_{N_R} \lesssim 0.1\, M_{W^\prime}$. Here, the RHN will be produced with a large boost in the transverse plane. Therefore, the decay of the RHN (i.e., $N_R\to \ell W^{\prime*}\to \ell jj$) will be highly collimated and produce a fatjet that can be used for reconstructing the $N_R$ wholly or at least, partially. This will give rise to one lepton and a $N_R$ jet ($J_{N_R}$). If the RHN is long lived, it will lead to a displaced vertex signature; i.e., its decay length roughly would lie in the range $\lt[10^{-3}-1\rt]$ m. In this case, the merged $J_{N_R}$ appears at a distance visibly away from the primary vertex. If $N_R$ decays outside the detector, it produces the invisible signature. 

\end{enumerate}  

As mentioned earlier, unlike the KS signal, our signal would have no same-sign lepton pair in the final state because of the small keV-scale $\mu$. 
However, our signal is also kinematically different from the KS signal even though we also consider the $M_{W^\prime} > M_{N_R}$ kinematic region since the RHN decays through an on-shell $W$ boson in our case [Fig.~\ref{fig:Signal}].
More specifically, a $W^\prime$ boson, produced from the $pp$ collision, decays into a $\m^-$ or a $\m^+$ and the second-generation RHN, $N^2_R$ (we choose the second generation because muons have better identification efficiency at the LHC than  electrons and taus.) Since we assume the other RHNs, i.e., $N^1_R$ and $N^3_R$, are heavier than $W^\prime$, it cannot decay to these RHNs.  
The $N_R^2$ then decays to a charged lepton and a boosted $W$ boson, which then decays hadronically. 
Thus, the final-state particles include two opposite-sign leptons and a $W$-like fatjet. The presence of a $W$-like fatjet is then a distinguishing feature of our signal. Beyond these, there are other distinguishing features of our signal, as 
seen from the distributions shown in Fig.~\ref{fig:dist}. Even if one ignores the absence of the same-sign lepton pair in the final state, these features can be used to discriminate between these two topologies if the experiments observe a significant number of $\ell\ell j j$ signal events.

\section{Existing $W^\prime$ searches and bounds}
\label{sec:exclu}

\noindent
We briefly review here the recent LHC direct search limits on $W^\prime$. In experimental searches, decays of $W^\prime$ to different fermionic modes ($\ell\nu$, $jj$, $tb$ and $N\ell$) and diboson modes ($WZ$ and $WH$) are considered; in our model, the $W\leftrightarrow W^\prime$ mixing suppresses the $W'\to \ell\nu,WZ,WH$ mode whereas the other decay modes, controlled by $g_R$, are not mixing angle suppressed.
\medskip

\noindent
\textbf{Searches for the KS process:} Recently, the CMS Collaboration has performed a search for the KS process  in the final states containing a pair of same-flavour charged leptons ($e$ or $\m$, with the same or opposite electric charges) and two jets~\cite{CMS:2021qef} at $\sqrt{s} = 13$ TeV with $137$ fb$^{-1}$ integrated luminosity. Assuming $g_L = g_R$, the search excludes $W'$ with mass up to $\sim 5$~TeV  with $95$\% confidence level (CL). Earlier, the ATLAS Collaboration has searched for the resolved~\cite{ATLAS:2018dcj} and merged~\cite{ATLAS:2019isd} topologies and obtained similar exclusion limits. Reference~\cite{CMS:2018iye} puts a lower limit of $3.5$ TeV on the mass of $W^\prime$ (assuming the mass of the third-generation RHN to be the half of $M_{W^\prime}$) from the search for the KS process in the $\tau\tau jj$ final state. (A slightly different process where the $W'$'s in the KS process are replaced by $W$'s is considered in Ref.~\cite{CMS:2018jxx}. This process is sensitive to the Yukawa couplings involved in the $N_R\to \ell W$ decay. The same process can also lead to displaced vertex signature if the decay width of $N_R$ is small~\cite{ATLAS:2019kpx,CMS:2021lzm}.)
\medskip

\begin{figure*}
\centering
\captionsetup[subfigure]{labelformat=empty}
\subfloat[(a)]{\includegraphics[width=0.4\textwidth]{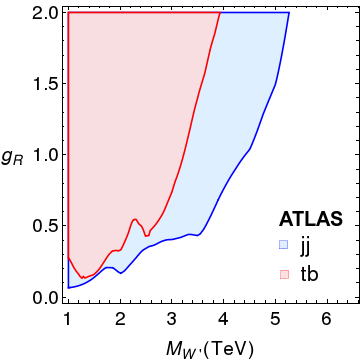}\label{fig:ATLASexclu}}\hspace{1cm}
\subfloat[(b)]{\includegraphics[width=0.4\textwidth]{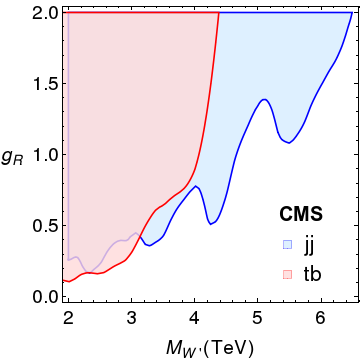}\label{fig:CMSexclu}}\\
\caption{Exclusion regions in the $M_{W'}$-$g_R$ plane obtained by recasting the dijet and $tb$ resonance search results by ATLAS and CMS.}\label{fig:brazil}
\end{figure*}

\noindent
\textbf{Searches for a dijet resonance:} Both ATLAS and CMS have searched for a $jj$ resonance~\cite{Aad:2019hjw,Sirunyan:2019vgj} at the $13$ TeV LHC with $139$ and $137$ fb$^{-1}$ of integrated luminosities, respectively. The ATLAS search rules out 
a sequential $W'$ with  $M_{W^\prime}\lesssim 4$ TeV, and the CMS study rules out $M_{W^\prime}\lesssim 3.6$ TeV.
 We have recast the observed limits from these two searches to obtain bounds on $g_R$,  as shown in Figs.~\ref{fig:ATLASexclu} and~\ref{fig:CMSexclu}. In the ATLAS search recast, we have appropriately factored in the variation of detector acceptance ($A$) with $M_{W^\prime}$. However, we have assumed a flat $A=0.5$ for the CMS search recast.
\medskip

\noindent
\textbf{Searches for a $tb$ resonance:}
The ATLAS Collaboration has presented a combined exclusion limit for the $W^\prime$ decaying through the $tb$ final state, with hadronic~\cite{ATLAS:2018uca} and leptonic~\cite{ATLAS:2018wmg} top decays  with $36.1$ fb$^{-1}$ integrated luminosity at the $13$ TeV LHC.  For a sequential $W'$ model, $M_{W'}\lesssim 3.15$ TeV has been ruled out at $95$\% CL. The CMS Collaboration also has performed the search
for a $W^\prime$ decaying into a $tb$ pair, in the all-hadronic mode using the $13$ TeV LHC data with $137$ fb$^{-1}$ of integrated luminosity~\cite{Sirunyan:2021sbg}.  The CMS search excludes the $W^\prime$ masses below $3.4$ TeV. We recast these limits as well [see Figs.~\ref{fig:ATLASexclu} and~\ref{fig:CMSexclu}]. 

\begin{table}[t]
\centering{
\begin{tabular}{lcc}
\hline
Experiment                                      & Luminosity (fb$^{-1}$)& Observed limit (TeV)\\ \hline
ATLAS dijet~\cite{Aad:2019hjw}                 &$139$& $3.80$ \\
CMS dijet~\cite{Sirunyan:2019vgj}              &$137$& $3.60$ \\
ATLAS $tb$~\cite{ATLAS:2018wmg}                &$36.1$& $3.45$ \\
CMS $tb$~\cite{Sirunyan:2021sbg}               &$137$& $3.50$  \\\hline
\end{tabular}}\label{tab:limits}
\caption{Summary of the $95\%$ CL exclusion limits on $W^\prime$ obtained from recasting the LHC experiments (assuming only one RHN decay mode of $W'$ is open).}
\end{table}

We summarise the dijet and $tb$ resonance limits obtained after recasting the searches, under the assumption that only one RHN decay mode is open, in Table~\ref{tab:limits}.
\medskip

\noindent
\textbf{Other searches:}
There are other searches for $W^\prime$ in various decay channels. For example, the charged lepton + missing transverse energy channel \cite{Aad:2019wvl,CMS:2021pmn,Aaboud:2018vgh,CMS:2018fza}, the $WH$ channel \cite{ATLAS:2021pqe,ATLAS:2020qiz,CMS:2021lyi,CMS:2019qem}, the $WZ$ channel \cite{ATLAS:2019nat,ATLAS:2020fry,CMS:2021lyi,CMS:2019qem}, etc. However, all these decays occur through $W$-$W^\prime$ mixing, which is small in the LRSM. Hence, these searches do not constrain the parameter space of our model. 

In principle, as demonstrated in Refs.~\cite{Mandal:2015vfa,Mandal:2016csb,Mandal:2018kau,Bhaskar:2021pml}, one can also recast other searches in the $\ell\ell j j$ channel (e.g., the leptoquark searches) to obtain bounds. Here, however, we ignore such bounds on $W^\prime$, as such limits are expected to be weaker than the direct ones.

\section{RHN decays through a $W$ boson}
\label{sec:lhcpheno}
\noindent
We implement the Lagrangian terms relevant for the productions and decays of $W'$ and $N_R$ in \textsc{FeynRules}~\cite{Alloul:2013bka}, and obtain the UFO~\cite{Degrande:2011ua} model files. We use the \textsc{NNPDF2.3} parton distribution functions to generate signal and the SM background events in \textsc{MadGraph5}~\cite{Alwall:2014hca}. The generated events are passed through \textsc{Pythia8}~\cite{Sjostrand:2014zea} for showering and hadronisation to \textsc{Delphes}~\cite{deFavereau:2013fsa} for detector simulation. We use the anti-$k_t$ jet clustering algorithm~\cite{Cacciari:2008gp} in \textsc{FastJet}~\cite{Cacciari:2011ma} to cluster jets from the tower objects. We use two types of jets in our analysis, namely, AK4-jets with jet radius parameter $R=0.4$ and AK8-fatjets with $R=0.8$~\cite{Bhaskar:2021gsy}. In this paper, we use the symbol $j$ for AK4-jets and $J$ for AK8-fatjets. We tag $b$ jets from the AK4-jets.

The process of our interest is
\begin{equation}
    pp\to (W^\prime)^{\pm} \to N_R \ell^{\pm} \to (W_h^{\pm}\ell^{\mp})\ell^{\pm}
\end{equation}
where, as mentioned before, $W_h^{\pm}$ denotes a hadronically decaying $W$ boson. Depending on the masses of the $W^\prime$ and $N_R$, the $W_h$ can be sufficiently boosted and form a two-pronged fatjet. We employ jet-substructure 
techniques to tag a boosted $W_h$ with high efficiency. Because of the pseudo-Dirac nature of the $N_R$, we only have opposite-sign dilepton accompanied by a boosted $W$ jet in the final state in our case. Leptons originating in the decays of a TeV-scale $W^\prime$ or $N_R$ will also have high transverse momenta ($p_T$). Therefore, the signature our signal would be two high-$p_T$ same flavor-opposite-sign leptons and a $W$-like two-pronged fatjet. Since there is no missing energy, our signal channel is fully reconstructible, in principle. 

There is a possibility of the displaced vertex appearing if the RHNs are long lived. This can happen if the decay couplings of the RHNs are very small. In our model, the Yukawa couplings (which control the decay of RHNs) are in the range $\sim 0.01-0.1$. We would not have a displaced vertex for a TeV-scale particle for these values. Moreover, for a heavy particle, its decay length would not be enhanced due to the time dilation effect. 

\begin{figure*}
\captionsetup[subfigure]{labelformat=empty}
\subfloat[(a)]{\includegraphics[height=7cm,width=7.5cm]{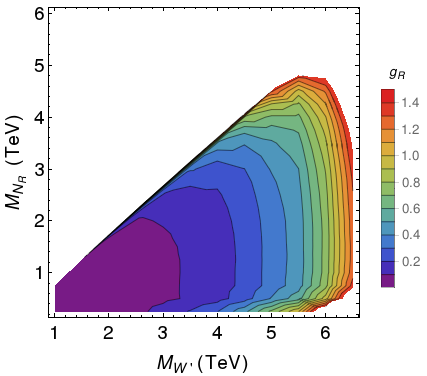}}\hspace{1cm}
\subfloat[(b)]{\includegraphics[height=7cm,width=7.5cm]{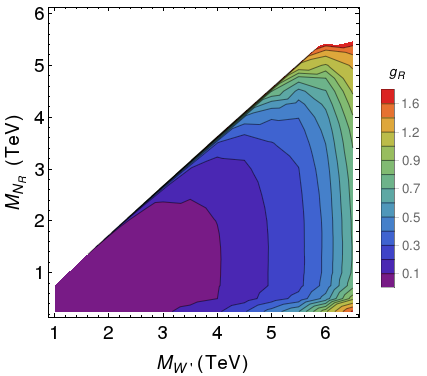}}
\caption{The regions in the $M_{W'}$-$M_{N_R}$ plane that can be (a) discovered with $5\sg$ significance and (b) excluded with $2\sg$ significance at the HL-LHC. The contours are for different $g_R$ values. In the left (right) plot, the regions with same colour can be discovered (excluded) with $5\sg$ ($2\sg$) significance or more.}\label{fig:MWpMNR}
\end{figure*}

\subsection{Background processes}
\noindent
The following SM processes with large cross sections form the relevant background of our signal:
\begin{itemize}
\item $Z+jets$: This process forms the dominant background. 
We generate it by simulating the $p p \to Z/\gamma \to \ell\ell$ process matched up to two extra partons. Here, the two high-$p_T$ leptons can arise from the leptonic decays of the $Z$ boson, and the QCD jets can be misidentified as the $W$-like fatjet. Since the invariant mass of the two leptons peaks at the $Z$ mass, this background is controlled by a $Z$-mass veto.

\item $tt+jets$: The SM top pair production can also provide us two high-$p_T$ leptons when both the tops decay leptonically. Additionally, a $W$-like jet can come from the QCD jets. The contribution of this process in the background is significant in our case. We generate the $t\bar t$ events by matching the parton showers (PSs) with up to two additional jets.

\item $tW+jets$: The SM $pp \rightarrow tW$ process contains two leptons in the final state when both the top quark and the $W$ boson decay leptonically. This process also contributes significantly to the background of our signal. In this case as well, the $W$-like jet arises from the QCD jets. We generate this process by jet-PS matching up to two extra jets. 

\item $VV+jets$: Here, $V$ denotes a $W$ or a $Z$ boson. There are four types of diboson processes, viz., $W_\ell W_\ell$, $W_h Z_\ell$, $Z_\ell Z_h$, and $Z_\ell H_h$ (the subscripts $\ell$ and $h$ represent leptonic and hadronic decay modes) that can act as sources of two high-$p_T$ leptons. In these cases, the $W$-like jet arises from the hadronic decay of a $V$ or QCD jets. Processes containing leptonically decaying $Z$ can be tamed by applying the $Z$-mass veto on the invariant mass of the lepton pair. Among all the diboson processes, the $W_\ell W_\ell$ process contributes maximally. We generate matched event samples (including up to two extra jets) of these processes. The total diboson contribution, however, is negligible after all the cuts.
    
\item $ttV$: The SM processes producing a top pair and a vector boson can act as backgrounds for our signal. We consider four cases, viz., $t_\ell t_\ell Z_h$, $t_ht_hZ_\ell$, $t_\ell t_\ell W_h$, and $t_\ell t_h W_\ell$, depending on the decays of the tops and vector boson. We generate these event samples without adding extra jets in the final state. Just like the diboson background, this background, too, contributes negligibly to the total background after the cuts.

\item $W+jets$: This process can contribute in the background when the $W$ decays leptonically, and a jet is misidentified as a lepton. It is one of the major background sources for the same-sign dilepton signature. However, in the opposite-sign dilepton case, its contribution to the total background is small since the efficiency of jet faking as a lepton is very small, $\sim 10^{-4}$~\cite{Curtin:2013zua}. Nevertheless, we consider the process here since its cross section is quite large ($\sim 10^5$ pb). We generate the process by jet-PS matching up to three additional jets. 
\end{itemize}

\begin{table}[!t]
\centering{\linespread{2}
\begin{tabular*}{\columnwidth}{l @{\extracolsep{\fill}} crc }
\hline
\multicolumn{2}{l}{Background } & $\sg$ & QCD\\ 
\multicolumn{2}{l}{processes}&(pb)&order\\\hline\hline
$V +$ jets~ \cite{Catani:2009sm,Balossini:2009sa}   & $Z +$ jets  &  $6.33 \times 10^4$& NNLO \\ \hline
$tt$~\cite{Muselli:2015kba}  & $tt +$ jets  & $988.57$ & N$^3$LO\\ \hline
Single $t$~\cite{Kidonakis:2015nna}  & $tW$  &  $83.10$ & N$^2$LO \\  \hline
\multirow{3}{*}{$VV +$ jets~\cite{Campbell:2011bn}}   & $WW +$ jets  & $124.31$& NLO\\ 
                  & $WZ +$ jets  & $51.82$ & NLO\\ 
                   & $ZZ +$ jets  &  $17.72$ & NLO\\ \cline{1-4}
\multirow{2}{*}{$ttV$~\cite{Kulesza:2018tqz}} & $ttZ$  &  $1.05$ &NLO+NNLL \\ 
                   & $ttW$  & $0.65$& NLO+NNLL \\ \hline
\end{tabular*}}
\caption{Total cross sections without any cut for the SM background processes considered in our analysis. The higher-order QCD cross sections are taken from the literature and the corresponding orders are shown in the last column. We use these cross sections to compute the $K$ factors which we multiply with the LO cross sections to include higher-order effects.}
\label{tab:BGxsec}
\end{table}
\begin{table*}[]
\centering{\linespread{3}
\begin{tabular*}{\textwidth}{l @{\extracolsep{\fill}} r r r r r r r}
\hline
Selection cut & Signal & $Z+\textrm{jets}$ & $tt+\textrm{jets}$ & $tW+\textrm{jets}$ & $WW+\textrm{jets}$ & $ttW$ & $ttZ$ \\
\hline\hline
Generation level (including $K$ factors) & 365 & $3.2\times 10^6$ & $7.3\times 10^5$ & $4.7\times 10^4$ & $3.9\times 10^4$ & 1128 & 403 \\
Number of muons $=2$ (any charge) & 256 & $2.5\times 10^6$ & $4.6\times 10^5$ & $3.3\times 10^4$ & $3.0\times 10^4$ & 673 & 240 \\
Number of $b$ jets~$=0$ (AK4 jets)& 254 & $2.5\times 10^6$ & $3.3\times 10^5$ & $3.3\times 10^4$ & $3.0\times 10^4$ & 468 & 167 \\
$p_T(\mu_1) > 300$ GeV, $p_T(\mu_2) > 100$ GeV & 253 & $1.0\times 10^5$ & $1.3\times 10^4$ & 2291 & 3988 & 53 & 19 \\
$M(\mu_1,\mu_2) > 200$ GeV & 251 & $9.8\times 10^4$ & $1.3\times 10^4$ & 2274 & 3939 & 52 & 19 \\
Number of fatjets~$\geq 1$ (AK8 jets) & 243 & $3.7\times 10^4$ & 9136 & 1432 & 1758  & 45 & 16 \\
$p_T(J_1) > 200$ GeV and $|\eta(J_1)|< 2.5$ & 222 & $2.1\times 10^4$ & 4124 & 584 & 1031  & 29 & 10 \\
$S_T > 1500$ GeV & 207 & 3050 & 556 & 47 & 178 & 2 & 1 \\
$M(J_1,\mu_1,\mu_2) > 0.8\times M_{W'}$ & 199  & 431 & 53 & 6  & 23 & $<1$ & $< 1$ \\
\hline  
\end{tabular*}}
\caption{Number of signal and background events obtained after the selection cuts at the $\sqrt{s}=14$~TeV LHC with $\mc{L}=3000$~fb$^{-1}$. The signal events are obtained for the benchmark parameters $M_{W'}=3$ TeV and $M_{N_R}=1$ TeV with $g_R=0.1$. For better statistics, we apply strong generation-level cuts (defined in the text) during event generation.}\label{tab:cutflow}
\end{table*}

We generate all the background processes discussed above at the leading order with \texttt{MadGraph5}. The relevant background processes and their cross sections (at the highest order in QCD available in the literature) are listed in Table~\ref{tab:BGxsec}. From the cross sections, we compute the $K$ factors to incorporate the higher-order effects in our analysis. Before cuts, some of the background processes are large. But since our signal belongs to a specific region of the phase space, we generate all the background processes with some strong generation level cuts to save computation time. Technically, this might lead to a small bias in the event samples. We, however, we ignore it for simplicity. The generation level cuts which we use are:
\begin{itemize}
    \item Transverse momentum $(p_T)$ on the leptons: $p_T(\ell_1)$, $p_T(\ell_2) > 100$~GeV.
    \item Invariant mass of the lepton pair $M(\ell_1,\ell_2) > 120$~GeV. 
\end{itemize}
The leptons are ordered according to their $p_T$. The cut on  $M(\ell_1,\ell_2)$ is applied to reduce the background involving $Z\to\ell\ell$ decay (the $Z$-mass veto). 

\subsection{Signal selection}
\noindent
For the final selection, we have divided the signal into low-mass ($M_{W^\prime}<3$~TeV) and high-mass ($M_{W^\prime}>3$~TeV) regions. As explained before, we demand two opposite-sign same flavour leptons and at least one $W$-like AK8-fatjet. Additionally, we also demand there is no $b$-tagged AK4 jet in the final state. The following selection cuts have been used for the two regions:\\

\noindent
\underline{Low-mass region ($M_{W^\prime}<3$~TeV):}
\begin{enumerate}
\item Transverse momentum on leptons, $p_T(\ell_1) > 300$~GeV, $p_T(\ell_2) > 100$~GeV.
\item Invariant mass of the lepton pair, $M(\ell_1,\ell_2) > 200$~GeV.
\item Scalar sum of $p_T$ of all visible objects, $S_T > 0.6\times M_{W^\prime}$.
\item Missing transverse energy, $\slashed{E}_T < 100$~GeV.
\item Mass of the leading AK8-fatjet, $|M(J_1) - M_W| < 20$~GeV.
\item $N$-subjettiness ratio of the leading AK8-fatjet, $\tau_{21}(J_1) < 0.35$.
\item Invariant mass on the leading AK8-fatjet and dilepton system, $|M(J_1,\ell_1,\ell_2) - M_{W^\prime}| < 200$~GeV.
\end{enumerate}
\medskip

\noindent
\underline{High-mass Region ($M_{W^\prime}\geq 3$~TeV):}
\begin{enumerate}
\item Cut 1 and cut 2 of the low mass region.
\item Scalar sum of the transverse $p_T$ of all visible objects, $S_T > 1500$~GeV
\item Invariant mass on the leading fatjet and dilepton system $M(J_1,\ell_1,\ell_2) > 0.8\times M_{W^\prime}$ (for $M_{W^\prime} < 5.5$~TeV) and $M(J_1,\ell_1,\ell_2) >  4400$~GeV (for $M_{W^\prime} \geq 5.5$~TeV)
\end{enumerate}

\begin{table*}[]
\centering{\linespread{3}
\begin{tabular*}{\textwidth}{l @{\extracolsep{\fill}} r l r l r l r }
\hline 
Variable & Importance & Variable & Importance & Variable & Importance & Variable & Importance \\ 
\hline \hline
$p_T(J_1)$   & $1.0 \times 10^{-1}$ & $M(J_1)$ & $9.6 \times 10^{-2}$ & $M(\mu_1,\mu_2)$ & $1.3 \times 10^{-1}$ & $\Dl R(J_1,\mu_1)$ & $6.0 \times 10^{-2}$ \\ 
$p_T(\mu_1)$ & $6.4 \times 10^{-2}$ & $M(J_1,\mu_1)$ & $6.4 \times 10^{-2}$ & $M(J_1,\mu_1,\mu_2)$ & $6.5 \times 10^{-2}$ & $\Dl R(J_1,\mu_2)$ & $6.5 \times 10^{-2}$ \\ 
$p_T(\mu_2)$  & $9.9 \times 10^{-2}$ & $M(J_1,\mu_2)$ & $1.2 \times 10^{-1}$ & $\tau_{21}(J_1)$ & $8.4 \times 10^{-2}$ & $\Dl R(\mu_1,\mu_2)$ & $5.3 \times 10^{-2}$ \\ 
\hline 
\end{tabular*} }
\caption{Input variables used in the multivariate analysis to separate the signal and the KS-like process and their relative importance.}
\label{tab:MVAvar}
\end{table*}

\begin{figure*}[t]
\centering
\captionsetup[subfigure]{labelformat=empty}
\subfloat[]{\includegraphics[width=0.9\textwidth]{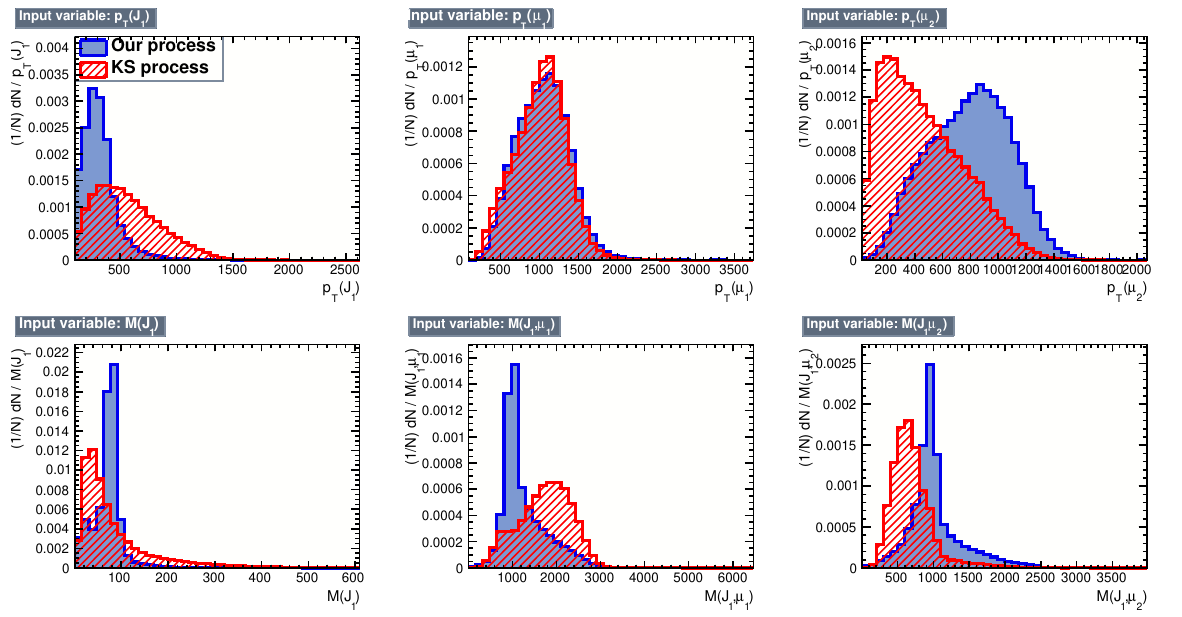}\label{fig:var1}}\\
\subfloat[]{\includegraphics[width=0.9\textwidth]{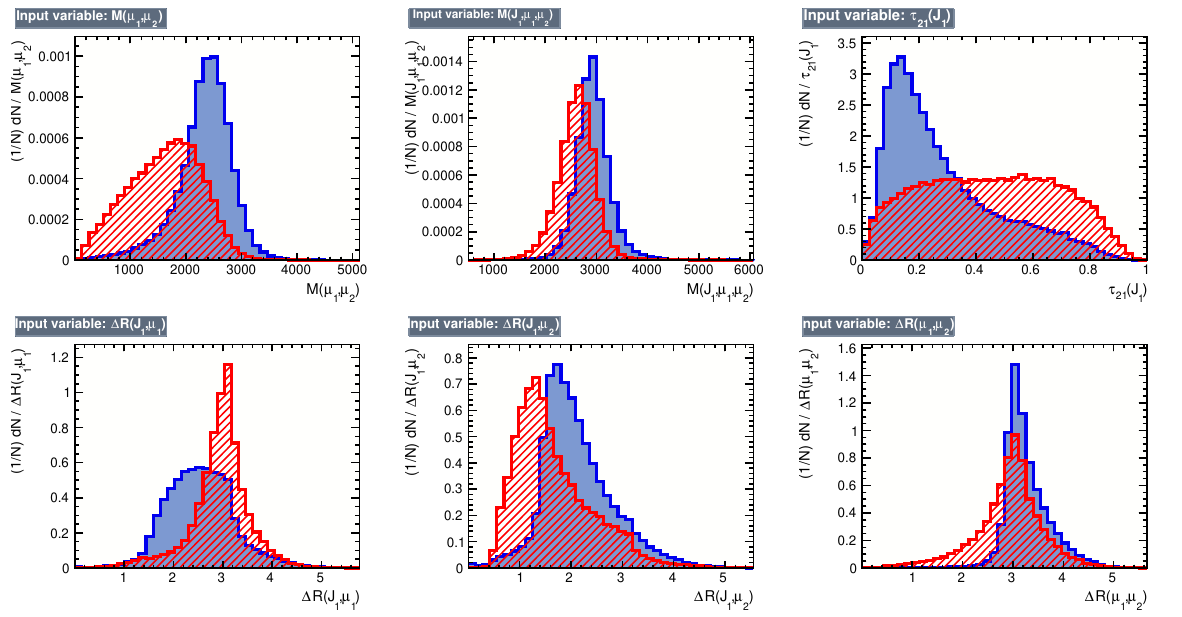}\label{fig:var2}}
\caption{Distributions of the input variables used in the multivariate analysis to separate our process (blue) from the KS-like one (red) for the benchmark point $M_{W^{'}}=3$~TeV and $M_{N_R}=1$~TeV with $g_R=0.1$.}\label{fig:kinvar}
\end{figure*}

\begin{figure}[t]
\includegraphics[width=0.94\columnwidth]{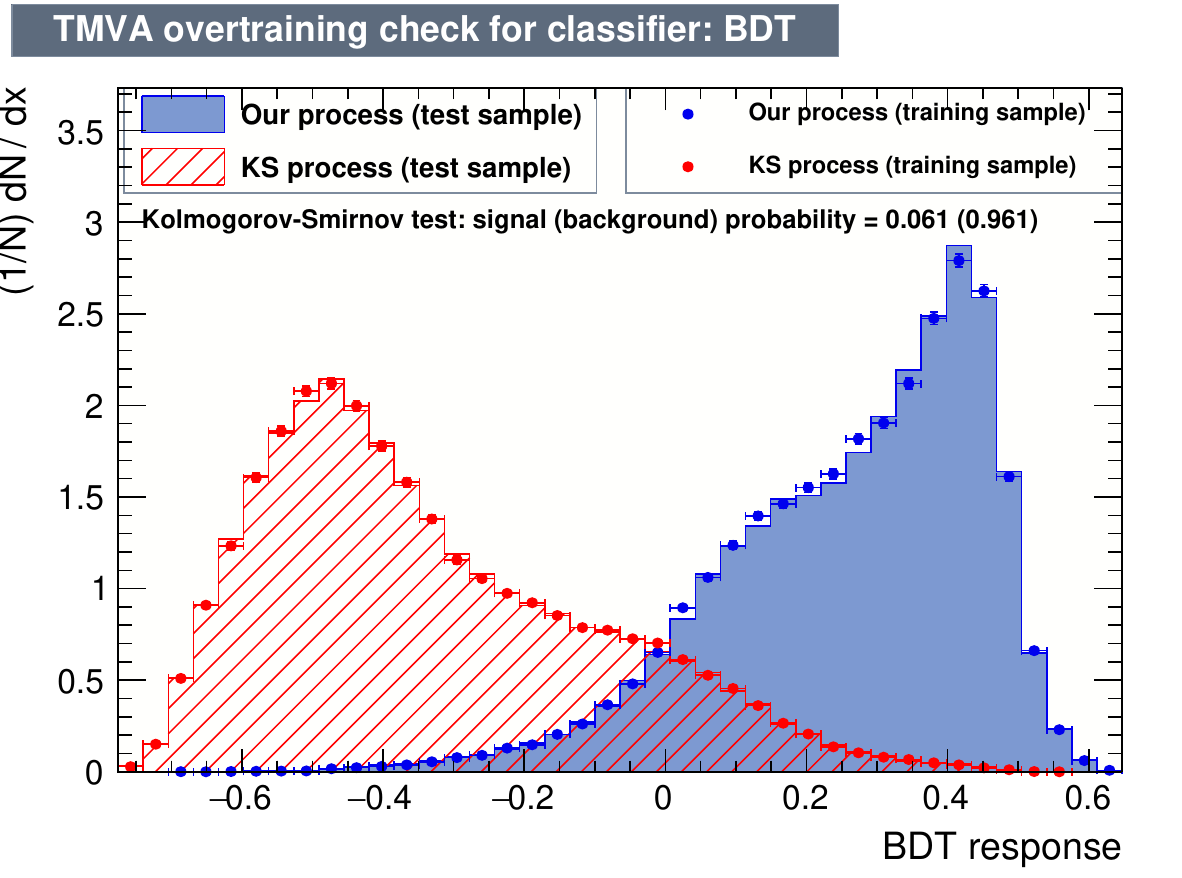}
\caption{BDT response for the benchmark point $M_{W^{'}}=3$~TeV and $M_{N_R}=1$~TeV with $g_R=0.1$.}\label{fig:BDT}
\end{figure} 

We show the effect of these cuts on the signal and the background processes for the high-mass benchmark point $M_{W'}=3$ TeV and $M_{N_R}=1$ TeV with $g_R=0.1$ in Table~\ref{tab:cutflow}. We optimize the cuts so that about $50$\% signal events are retained but the background is reduced by $4$ orders of magnitude.

We follow a simple procedure to estimate the $W+jets$ contribution to the background. To reduce the computation time, we generate it by applying a cut, $p_T(\ell)>250$ GeV. We assume the second lepton arises from a faking (AK4) jet. When a jet is declared a lepton, we ensure it is separated from the AK8-fatjet by $\Dl R > 0.8$. All the other cuts remain the same as above. With a jet-faking efficiency of $10^{-4}$~\cite{Curtin:2013zua}, less than two events survive the cuts at the HL-LHC for the benchmark point $M_{W'}=3$ TeV and $M_{N_R}=1$ TeV. Even if we assume the efficiency to be one
order higher for a conservative estimate, only $10-12$ events survive at the end, making this background unimportant.

\subsection{Signal significance}
\noindent
After applying the above cuts, let the number of surviving signal and background events at a given luminosity be denoted by $N_S$ and $N_B$, respectively. From these, one can get an estimation of the statistical significance of the signal from the formula below:
\begin{equation}
\mc{Z} = \sqrt{2(N_S+N_B)\ln\lt(\dfrac{N_S+N_B}{N_B}\rt) - 2N_S}.
\end{equation}
We estimate $\mc{Z}$ for the HL-LHC, i.e., for $3$~ab$^{-1}$ integrated luminosity and $\sqrt{s}=14$~TeV. Figure~\ref{fig:MWpMNR} shows the projected discovery  ($5\sg$ significance) and $2\sg$ exclusion regions in the $M_{W'}$-$M_{N_R}$ plane. Our signal process mainly depends on three model parameters.
Among them, $M_{W'}$ and $M_{N_R}$ are kinematic in nature; i.e., they can affect the signal distributions. The free gauge coupling $g_R$, on the other hand, is nonkinematic and only scales the distributions. (This is true as long as the narrow-width approximation is valid. Large width can affect various distributions and can potentially change the reach. However, we ignore the large width effects for simplicity). 
In Fig.~\ref{fig:MWpMNR}, the contours correspond to fixed $g_R$ values to achieve $5\sg$ discovery significance, i.e., a region with a particular colour will have signal significance $5\sg$ or more for the corresponding value of $g_R$. There are two special regions in the plot where the sensitivity is low. One is with $M_{N_R} \lesssim M_{W'}$ where the BR($W'\to N_R\ell$) is phase-space suppressed. It can be checked from Fig.~\ref{fig:BRWpNR}. The other region is where $M_{N_R}\lesssim 0.1\times M_{W'}$. This is the merged region for which a different analysis strategy is required~\cite{Mitra:2016kov}.

\section{Summary and conclusions}
\label{sec:sumcon}
\noindent
In this paper, we have studied a hitherto experimentally unexplored signature of the left-right symmetric models with the inverse seesaw mechanism for neutrino mass generation. In particular, we have considered a channel where sequential decays of two TeV-scale new particles; a heavy charged gauge boson, $W'$; and a pseudo-Dirac right-handed neutrino, $N_R$, lead to a final state with two high-$p_T$ same-flavour-opposite-sign leptons and a $W$-like fatjet.

A similar process involving a $W'$ and a Majorana $N_R$, known as the Keung-Senjanovi\'{c} process, has been already searched for at the LHC as a test of lepton-number violation. The final state of the KS process is a same-sign lepton pair and a pair of jets. It originates from Drell-Yan production of a $W'$. The $W^\prime$ first decays to a $N_R\ell$ pair, and then the $N_R$ undergoes through a three-body decay, $N_R\to \ell jj$, via an off-shell $W'$. If the RHN is completely Majorana type, both the same-sign and opposite-sign dilepton final states will be present with equal rate, but if it is pseudo-Dirac type, the same-sign dilepton final state will be (almost) absent.
There is a possibility of two-body decays of $N_R\to \ell W\to \ell jj$ that gives the same $\ell\ell jj$ final state but through an on-shell $W$. However, in the standard type-I seesaw mechanism with TeV-scale RHNs, the partial widths of the two-body decay modes of $N_R$ are negligible due to the small Yukawa couplings involved in the decays. In the LRSM, a three-body decay of $N_R$ through off-shell $W'$ opens up, which, despite the phase-space suppression, can overcome the two-body decays to become the dominant decay mode of $N_R$. This leads to the KS process but at the expense of small Yukawa couplings required in the type-I seesaw mechanism to have a TeV-scale $N_R$ in the spectrum.

We invoke the inverse seesaw mechanism to have a natural TeV-scale $N_R$ with order-$1$ Yukawa couplings controlling the two-body decays. Consequently, the two-body decay BRs overcome the three-body one when the $N_R$ is embedded in the LRSM. In this setup, we lose the clean same-sign signature of the KS process, but a similar opposite-sign dilepton signature arises whose kinematic nature is very different from the KS process. In association with the lepton pair, we also have a pair of jets similar to the KS process in the final state. However, two jets are collimated in our process as they come from the decay of a boosted $W$ boson. Therefore, our final state contains two same-flavour-opposite-sign leptons and a $W$-like fatjet. We design a set of selection cuts using the jet-substructure variables to observe the signal over the large background with a significance of more than $5\sg$ at the HL-LHC. We have found that a $W'$ with $g_R\approx g_L$ and mass up to $\sim 6$~TeV can be discovered at the HL-LHC through this channel.

The large Yukawa couplings controlling the two-body decays of $N_R$ can, in principle, lead to substantial lepton-flavour violation~\cite{Cai:2017mow,FileviezPerez:2017zwm}. Such violation can be sizable in some regions of the parameter space in the neutrino sector. It will be interesting to study the allowed parameter space of our model by the lepton-flavour-violation data in the future. On the other hand, if the decay couplings involved in the RHN decay are very small, they can lead to displaced vertex signature, another interesting direction to investigate.

\begin{acknowledgments}
\noindent
A.M. acknowledges financial support from SERB-DST through Project No. EMR/2017/001434. M.T.A. is financially supported by the DST through the INSPIRE Faculty grant. T.M. acknowledges the use of high-performance computing facility at IISER-TVM.
\end{acknowledgments}

\appendix
\section{Discriminating signals---A simple multivariate analysis}

\noindent
As discussed earlier, the fatjet mass distribution is a very good discriminator of the two kinematically different topologies arising from the decay of RHN through an on-shell $W$ or an off-shell $W'$. We would expect a two-prong $W$-like fatjet when the $N_R$ decays through an on-shell $W$. Along with other kinematic distributions with good discrimination power, this feature can be used in a multivariate analysis (MVA) to distinguish these two different kinematic regions.

To illustrate this point, we perform a simple but indicative boosted decision tree (BDT-)based MVA. In particular, we use the adaptive BDT algorithm in the TMVA package~\cite{Hocker:2007ht}. In Fig.~\ref{fig:kinvar}, we show the distributions of $12$ variables used in the analysis. Their relative importance are shown in Table~\ref{tab:MVAvar}. From Fig.~\ref{fig:BDT}, we see that these two processes are well separated in the BDT response distribution. Roughly, a BDT cut around $\gtrsim 0$ is sufficient to discriminate between the two processes at the HL-LHC. 
\relax

\bibliography{LRSMISM}
\bibliographystyle{JHEPCust}

\end{document}